# A Study of EV BMS Cyber Security Based on Neural Network SOC Prediction

Syed Rahman, Haneen Aburub, Yemeserach Mekonnen, and Arif I. Sarwat
Department of Electrical and Computer Engineering
Florida International University
Miami, Florida
srahman9713@gmail.com, haburub@fiu.edu, ymeko001@fiu.edu, asarwat@fiu.edu

*Abstract*—Recent changes to greenhouse gas emission policies are catalyzing the electric vehicle (EV) market making it readily accessible to consumers. While there are challenges that arise with dense deployment of EVs, one of the major future concerns is cyber security threat. In this paper, cyber security threats in the form of tampering with EV battery's State of Charge (SOC) was explored. A Back Propagation (BP) Neural Network (NN) was trained and tested based on experimental data to estimate SOC of battery under normal operation and cyber-attack scenarios. NeuralWare software was used to run scenarios. Different statistic metrics of the predicted values were compared against the actual values of the specific battery tested to measure the stability and accuracy of the proposed BP network under different operating conditions. The results showed that BP NN was able to capture and detect the false entries due to a cyber-attack on its network.

## I. INTRODUCTION

Greenhouse gasses affect the environment negatively by tabbing the heat and making the planet warmer. A big reason for this is the burning of fossil fuels for the transportation sector. The transportation sector currently accounts for roughly a quarter of greenhouse gas emissions [1]. Current vehicle emission restrictions are not very strict, but some countries and even corporations have taken necessary measures to decrease global emissions. India has declared that by 2030 it will be illegal to sell fuel burning car and all car sold after 2030 will be powered by electricity[2]. Volvo has also declared that by 2019 they will only sell hybrid and fully EVs [3]. These major changes in policy and industry will drastically reduce greenhouse gas emissions. Electric vehicles emit 54 percent less carbon emissions than internal-combustion counterparts and by 2050 the reduction could be even up to 70 percent as renewable sources of energy start to power the grid[1]. Integrating large number of EVs into the future smart grid will face several challenges from the vehicle and grid sides [4]–[6] .Cyber security is considered as one of the major issues that might affect EVs performance if not addressed carefully. The cyber-attack on Tesla model S highlighted the need for more research and investigation on the cyber security capabilities of EVs [5]. Hackers were able to hack into the Tesla Model S by [1] entertainment system, and take control of the vehicle [5]. In addition, the hackers were able to stop the vehicle immediately, at less than 5 miles per hour or idle, by applying the emergency handbrake [5]. At higher speeds, the hackers turned off the engine and stopped the vehicle [5]. The hackers could also lock/unlock the car remotely, control the radio and touch screens, and open/close the trunk [5]. While Tesla claimed that they have developed a solution, the fact is that these cars are vulnerable to cyber security threats should be further studied. There were six vulnerabilities in the car's security system, but Tesla emphasized that it was after having physical access to the Model S [5]. Later, researchers said they took control of the car remotely after having physical access to the entertainment system [7]. In [8], an access to a vehicle was achieved wirelessly through an Android Phone connected to an Open Vehicle Monitoring System (OVMS) server. Based on [9], the biggest threat of all is the damage that can be caused to the vehicle engine since the hackers were able to gain access to the cars Electrical Control Unit (ECU). These alarming incidents prove that improving security of these systems is a must. EV battery management system (BMS) can be one of the hackers' targets to change the battery's settings, such as SOC values. BMS depends on SOC estimated values and some other values to make the battery work efficiently and extend its life. Therefore, a cyber-attack on the SOC estimation functionality and process would result in over- or under- estimation of the actual EV battery performance. NN-based SOC estimation has been used widely in the literature [7], [10]-[11]. However, to the best of our knowledge, none of the previous research studies covered the security part of SOC prediction in EV. This paper will implement Backpropagation (BP) for SOC estimation in EV, while considering the cyber threat aspect. The main contribution of this paper will be in ability of NN to detect a cyber-attack scenario on its SOC prediction functionality. This will be an alarming indicator of cyber-attack for the EV driver.

This work is supported by the National Science Foundation under grant No. 1553494. Any opinions, findings, and conclusions or recommendations expressed in this material are those of the authors and do not necessarily reflect the views of the National Science Foundation.

Section II presents the experimental set-up and tools used to collect and analyze data for BP training network. Section III provides a brief description about BP NN. In Section IV, the designed BP NN for SOC prediction is presented, and the results are discussed for SOC prediction under hacked and normal operating scenarios. Finally, Section V concludes the work and discusses possible future works.

## II. DATA COLLECTION AND ANALYSIS

Sealed Lead Acid (SLA) batteries were chosen for the test to represent an EV battery. The tested SLA batteries have 12V nominal voltage and 8000 mAh. Prior to testing, all batteries were fully charged, given 5 days of stability and their internal Open Circuit Voltage (OCV) were measured. The batteries were connected to a PCBA 5010-4 battery analyzer to control and monitor their current and voltage during charge and discharge cycle. The batteries temperature, OCV, capacity, and power were recorded for each cycle. The time durations of the charge and discharge cycles were controlled using the battery analyzer software. In this experiment, two different charging/discharging profiles were used; C/10 profile equivalent to 800mA and C/20 profile equivalent to 400mA. In addition, new and old batteries were tested. Two sets of the batteries #P009 & #P016 are new and #31393 is old. The old battery have been used in the field for 2 years and 10 months. Battery #P009 and #31393 were charged at a constant current of $0.1C_{10}$ A until the voltage of the batteries was increased to 14.7V, and charged at a constant voltage of 14.7V until the charging current fell to $0.01C_{10}$ A. After the charging step, batteries were discharged at a constant current of $0.1C_{10}$A until the voltage of the batteries reached to End of Discharge Voltage (EODV) of 10.2V. Battery #P016 were charged at a constant current of $0.05C_{10}$ A until the voltage of the batteries was increased to 14.7V, and charged at a constant voltage of 14.7V until the charging current fell to $0.01C_{10}$ A. After the charging step, batteries were discharged at a constant current of $0.05C_{10}$A until the voltage of the batteries reaches to EODV of 10.5V.

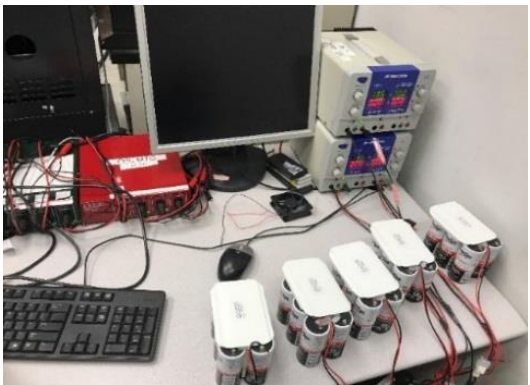

Fig. 1. SOC experimental setup

Section III will present the input and output experimental data used to study the effect of a cyber-attack on the NN structure designed for SOC prediction.

## III. NEURAL NETWORK

SOC estimation entails a complex nonlinear relationship between its inputs and outputs. In recent times, Artificial Neural Network (ANN) has shown promising applicability to non-linear heavy data-driven problems. ANN is an information processing system that mimic some of the performance characteristic of biological neural networks [12]. Generally, ANNs are classified in different category depending on the complexity of the problem at hand. They are distinguished from one another in terms of architecture, their activation functions in the processing elements and training method [13]. The BP NN learning algorithm is the most common NN learning algorithm currently in use that can provide suitable non-linear mapping and self-learning. BP implementation involves feedforward of input training pattern, computation of backpropagation error and adjustment of weights. Fig. 2 shows the basic general architecture of NN with backpropagation learning.

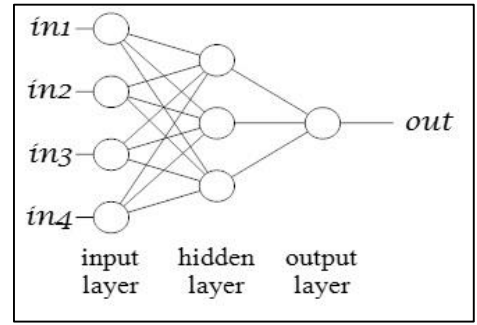

Fig. 2. Basic Multi-layer Feed Forward Network with back propagation learning.

Based on Fig. 2, an n number of neurons form a layer. Input layer is where the input data are fed in, the output layer provides the predicted output, and hidden layer connects the input and output layers [14]. Based on the experiment described in Section II.

## IV. RESULTS AND DISCUSSION

The proposed BP network model was implemented under different case scenarios; under normal operation and when the EV is under cyber-attack. The cycle number, average OCV, discharge current rate, end voltage, elapsed time, and test date are critical factors in SOC predication and therefore used as input element in the network. The output layer consist of one element which is SOC values. Fig. 3 shows the data of SLA battery #31393 that was used for training the NN. Battery #P009 was assumed to be the one under cyber threat. Fig, 4 shows the first 35 operating cycles of battery #P009 were used for testing the trained NN, and for predicting the #P009 battery SOC for the future cycles up to 76.

| Cycle Number | Avg. Voltage | Discharge Current | End Voltage | Avg. Temp | Elapsed Time H.M | Start Day | Start Month | Start Year | SOC (%) |
|---|---|---|---|---|---|---|---|---|---|
| 1 | 11.99 | 0.8 | 11.1 | 22.8 | 9.21 | 3 | 2 | 15 | 93.625 |
| 2 | 12.07 | 0.8 | 11.1 | 22.04 | 9.33 | 5 | 2 | 15 | 95.6 |
| 3 | 12.1 | 0.8 | 11.1 | 22.02 | 9.41 | 9 | 2 | 15 | 96.9 |
| 4 | 12.1 | 0.8 | 11.1 | 21.75 | 9.41 | 11 | 2 | 15 | 97.0 |
| 5 | 12.11 | 0.8 | 11.1 | 21.03 | 9.4 | 13 | 2 | 15 | 96.7 |
| 6 | 11.96 | 0.8 | 10.2 | 22.2 | 10.14 | 2 | 3 | 15 | 100.0 |
| 7 | 12 | 0.8 | 10.2 | 24 | 10.3 | 4 | 3 | 15 | 100.0 |
| 8 | 11.99 | 0.8 | 10.2 | 22.11 | 10.27 | 6 | 3 | 15 | 100.0 |
| 9 | 11.99 | 0.8 | 10.2 | 22 | 10.23 | 10 | 3 | 15 | 100.0 |
| 10 | 12 | 0.8 | 10.2 | 22 | 10.26 | 13 | 3 | 15 | 100.0 |
| 11 | 12 | 0.8 | 10.2 | 22.02 | 10.24 | 24 | 3 | 15 | 100.0 |
| 12 | 12.04 | 0.8 | 10.2 | 22 | 10.23 | 25 | 3 | 15 | 100.0 |
| 13 | 12 | 0.8 | 10.2 | 20 | 10.12 | 28 | 3 | 15 | 100.0 |
| 14 | 12.02 | 0.8 | 10.2 | 22.16 | 9.58 | 2 | 4 | 15 | 99.8 |
| 15 | 12.03 | 0.8 | 10.2 | 22.83 | 10.03 | 6 | 4 | 15 | 100.0 |
| 16 | 12.01 | 0.8 | 10.2 | 24 | 10.15 | 10 | 4 | 15 | 100.0 |
| 17 | 11.99 | 0.8 | 10.2 | 24 | 10.12 | 15 | 4 | 15 | 100.0 |
| 18 | 12.02 | 0.8 | 10.2 | 22 | 10.08 | 20 | 4 | 15 | 100.0 |
| 19 | 12.11 | 0.8 | 11.1 | 22.38 | 9.28 | 23 | 4 | 15 | 94.8 |
| 20 | 11.98 | 0.8 | 10.2 | 21.95 | 10.03 | 30 | 4 | 15 | 100.0 |

Fig. 3. A sample of battery #31393 data for training the NN

Root mean square (RMS) error was used as the statistical measure for accuracy of the NN. The best number of hidden layers was found by a trial and error process to find the best "minimum" RMS error. Fig. 5 shows the designed NN architecture for the purpose of predicting battery's #P009 SOC, and implemented on NeuralWare software. NeuralWare software has been used to compute the BP NN. NeuralWare is a software that develops and implements empirical modeling solutions based on neural networks [15]. It is mainly used for prediction, classification, or pattern recognition [15]. Based on Fig. 5, the number of neurons in the input layer is 9, and 1 in the output layer which represents the SOC. One hidden layer with 7 neurons was found to provide the best RMS value (10%). The activation function used is the Delta-Rule Sigmoid. Figs. 6 and 7 show the difference between the actual and predicted SOC values up to 40 and 60 cycles respectively.

| Cycle Number | Avg. Voltage | Discharge Current | End Voltage | Avg. Temp | Elapsed Time H.M | Start Day | Start Month | Start Year | SOC (%) |
|---|---|---|---|---|---|---|---|---|---|
| 1 | 12.03 | 0.8 | 11.09 | 21.76 | 9.33 | 16 | 2 | 15 | 95.59 |
| 2 | 12.08 | 0.8 | 11.09 | 22.74 | 9.31 | 17 | 2 | 15 | 95.24 |
| 3 | 11.99 | 0.8 | 11.09 | 20.13 | 9.31 | 20 | 2 | 15 | 93.08 |
| 4 | 12.07 | 0.8 | 11.09 | 23.81 | 9.21 | 23 | 2 | 15 | 93.54 |
| 5 | 12.09 | 0.8 | 10.19 | 22.74 | 9.16 | 24 | 2 | 15 | 92.73 |
| 6 | 11.99 | 0.8 | 10.19 | 24 | 9.53 | 26 | 2 | 15 | 98.91 |
| 7 | 11.98 | 0.8 | 10.19 | 22.72 | 9.42 | 27 | 2 | 15 | 97.13 |
| 8 | 11.98 | 0.8 | 10.2 | 24 | 9.38 | 3 | 3 | 15 | 96.48 |
| 9 | 11.99 | 0.8 | 10.19 | 24.06 | 9.35 | 5 | 3 | 15 | 95.84 |
| 10 | 11.99 | 0.8 | 10.19 | 22.16 | 9.25 | 6 | 3 | 15 | 94.25 |
| 11 | 11.97 | 0.8 | 10.19 | 23.57 | 9.19 | 10 | 3 | 15 | 93.26 |
| 12 | 11.98 | 0.8 | 10.19 | 24 | 9.21 | 16 | 3 | 15 | 93.60 |
| 13 | 12 | 0.8 | 10.19 | 25.3 | 9.16 | 18 | 3 | 15 | 92.74 |
| 14 | 12.01 | 0.8 | 10.19 | 24 | 9.07 | 20 | 3 | 15 | 91.28 |
| 15 | 12.01 | 0.8 | 10.19 | 22.97 | 8.59 | 24 | 3 | 15 | 89.99 |
| 16 | 12.02 | 0.8 | 10.19 | 24.15 | 8.51 | 26 | 3 | 15 | 88.64 |
| 17 | 12.02 | 0.8 | 10.19 | 20.14 | 8.37 | 28 | 3 | 15 | 86.23 |
| 18 | 12.02 | 0.8 | 10.19 | 22.69 | 8.39 | 30 | 3 | 15 | 86.55 |
| 19 | 12.03 | 0.8 | 10.19 | 22 | 8.31 | 1 | 4 | 15 | 85.24 |
| 20 | 12.06 | 0.8 | 10.19 | 23.59 | 8.13 | 3 | 4 | 15 | 82.30 |

Fig. 4. A sample of battery #P009 data for testing the NN

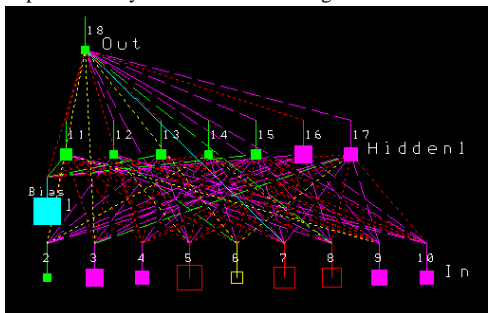

Fig. 5. BP NN architecture for SOC prediction

Equation 1 shows the mean absolute percentage error (MAPE) statistical metric that was used to measure the accuracy of the predicted values compared to the actual values [16].

$$MAPE = \frac{100}{N} \sum_{1}^{N} \left|\frac{A - P}{A}\right| \qquad (1)$$

Where $N$ is the number of cycles, $A$ is the actual value, and $P$ is the predicted value.

The backpropagation based NN provided good SOC prediction with 2% and 4% MAPE for the 40 and 60 cycle's cases respectively. Increasing the training data set and correlation between the batteries used for training and testing conditions could improve the prediction results.

A. Hacked SOC Prediction

In this section, two hacking scenarios were studied, in which the hacker could get access to the BMS SOC NN input and/or output layers, and replaced the battery #P009 testing data with the battery #P016 data set. The battery #P016 data was used as "hacking" data.

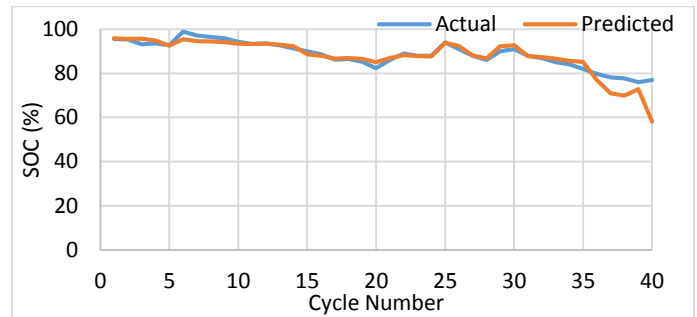

Fig. 6. Actual vs. predicted SOC values under normal EV BMS operating conditions up to 40 cycles.

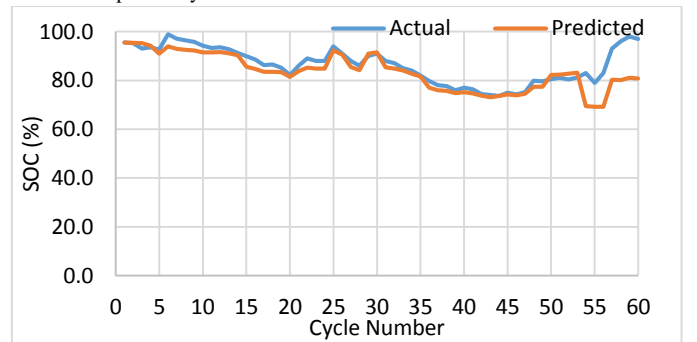

Fig. 7. Actual vs. predicted SOC values under normal EV BMS operating conditions up to 60 cycles.

1) *Case 1:* The first 35 operating cycles "testing data" of battery #P009 were replaced by the battery #P016 data in both the input and output layers of the designed NN architecture. Fig. 8 shows part of battery #P016 data.

Batteries #P016 and #P009 are two different types of new batteries and they were tested under two different test conditions, and provided different SOC values at each cycle as shown in Figs. 4 and 8. The designed NN architecture for the EV BMS SOC prediction was applied in this section. Figs. 9 and 10 provide a comparison between the actual and predicted SOC NN results for battery #P009 under the first hacking scenario for 40 and 60 cycles respectively. The MAPE values were approximately around 16% for both cycling cases. Compared to the normal operating conditions MAPE value, the MAPE of EV BMS SOC is higher under the hacking condition. The MAPE values were approximately around 16% for both cycling cases. Compared to the normal operating conditions MAPE value, the MAPE of EV BMS SOC is higher under the hacking condition. This indicates that the NN was able to detect tampering attempts by the hacker on the NN testing data. Figs. 11 and 12 clarify more this conclusion. Figs. 11 and 12 provide a comparison between the predicted EV BMS SOC values under normal and hacked conditions for 40 and 60 cycles respectively.

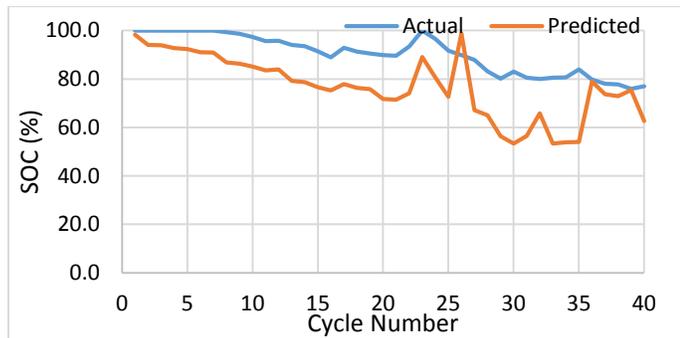

Fig. 8. A sample of battery #P016 data for hacking test of the NN

Equation 2 shows that the mean absolute percentage difference (MAPD) was used as a measure to interpret information from Figs. 11 and 12 since the comparison is between two values from the same category "predicted" [17].

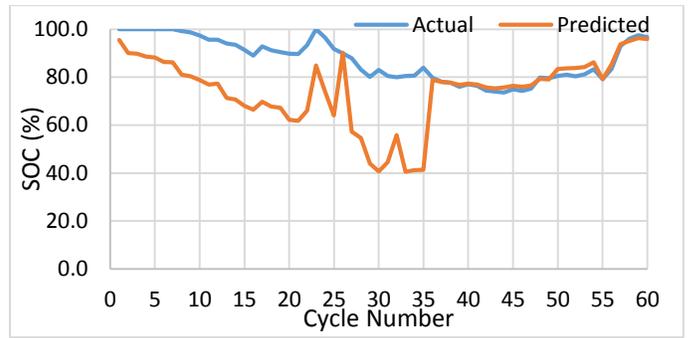

Fig. 9. Actual vs. predicted SOC values under case 1 hacked EV BMS operating conditions up to 40 cycles.

Fig. 10. Actual vs. predicted SOC values under case 1 hacked EV BMS operating conditions up to 60 cycles.

$$MAPD = \frac{100}{N}\sum_{1}^{N}\left|\frac{P_{normal} - P_{hacked}}{(P_{normal} + P_{hacked}/2)}\right| \quad (2)$$

MAPD were found to be 16% and 18% for the 40 and 60 operating cycles respectively.

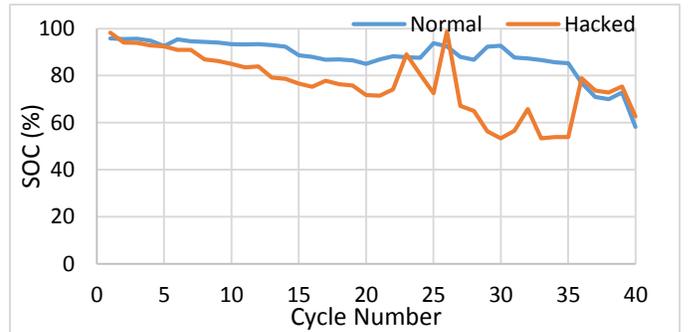

Fig. 11. Predicted EV BMS SOC values under normal and hacked operating conditions up to 40 cycles.

2) *Case 2:* The first 35 operating cycles "testing data" SOC values "output layer only" of battery #P009 were replaced by the battery #P016 data. Figs. 13 and 14 provide a comparison between the actual and predicted SOC NN results for battery #P009 under the second hacking scenario for 40 and 60 cycles respectively.

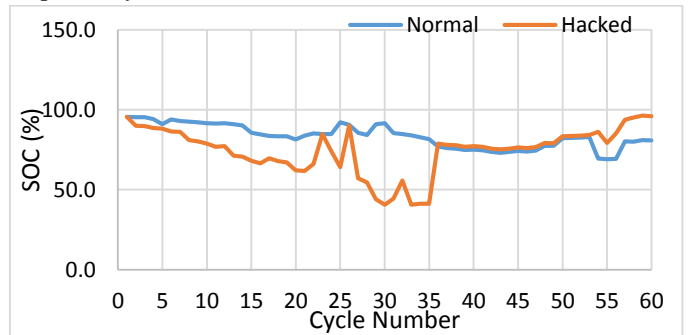

Fig. 12. Predicted EV BMS SOC values under normal and hacked operating conditions up to 40 cycles.

The MAPE was found to be 6%, which is a slightly higher value compared to normal operating condition value in part A. This shows that NN was still able to capture the small data tampering case.

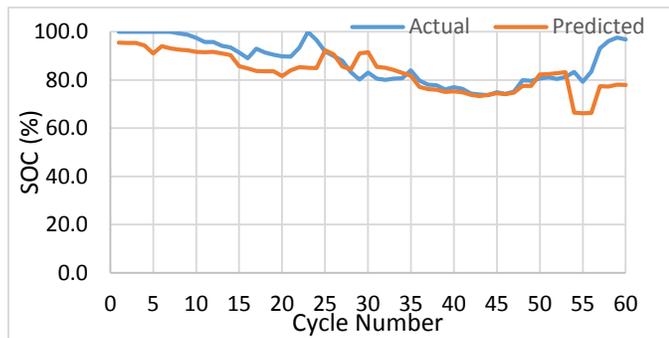

Fig. 13. Actual vs. predicted SOC values under case 2 hacked EV BMS operating conditions up to 60 cycles.

V. CONCLUSION AND FUTURE WORK

EV cyber security is one of the major issues that needs to be studied to ensure its stable and reliable operation in the future smart grid. This paper focused on one of the BMS security concerns, which is regarding SOC estimation. Many research papers in the literature have studied different techniques that can be used to predict the EV battery's SOC, such as NN. However, to the best of our knowledge, none of the previous research studies covered the security part of SOC prediction in EV. Experimental data from charge and discharge cycles of a lead-acid batteries have been used to train the NN. The NN was designed with 10% RMS. Different study cases have been applied on the NN to check its performance. NN was able to achieve 4% MAPE under normal operating conditions. Under a cyber-attack on the input and output layers of the NN, a 16% MAPE was achieved. While only attacking the output layer of the NN resulted in MAPE of 6%. The resulted MAPE values indicates that the NN was able to detect the corrupt numbers.

Future studies will focus on improving the NN security capabilities for attacks against SOC prediction functionality of EV. Other BMS and EV security aspects will be investigated as ell in future studies, such as brakes, steering/auto-pilot and locking systems.

5